\documentclass[twocolumn,amsmath,amssymb,prb,aps]{revtex4}%preprintnumbers,showpacs
\usepackage{graphicx}
\usepackage{amsmath}
\usepackage[perpage,symbol*]{footmisc}
\usepackage{epsfig}
\usepackage{epstopdf}

\begin{document}

%\begin{document}
\title{Engineering Photon Delocalization in a Rabi Dimer with a Dissipative Bath}
\date{\today}
\author{Fulu Zheng$^{1, 2}$, Yuyu Zhang$^{3}$, Lu Wang$^{2,4}$, Yadong Wei$^{1}$ and Yang Zhao$^{2}$\footnote{Electronic address:~\url{YZhao@ntu.edu.sg}}}

\address{$^{1}$ School of Physics and Energy, Shenzhen University, Shenzhen 518060, China}
\address{$^{2}$ School of Materials Science and Engineering,
Nanyang Technological University, Singapore 639798}
\address{$^3$ Department of Physics, Chongqing University, Chongqing 404100, China}
\address{$^4$ School of Science, Inner Mongolia University of Science and Technology, Baotou 041010, China}
\widetext

%%%%%%%%%%%%%%%%%%%%%%%%%%%%%%%%%%%%%%%%
\begin{abstract}
A Rabi dimer is used to model a recently reported circuit quantum electrodynamics system composed of two coupled transmission-line resonators with each coupled to one qubit. In this study, a phonon bath is adopted to mimic the multimode micromechanical resonators and is coupled to the qubits in the Rabi dimer. The dynamical behavior of the composite system is studied by the Dirac-Frenkel time-dependent variational principle combined with the multiple Davydov D$_{2}$ ans\"{a}tze. Initially all the photons are pumped into the left resonator, and the two qubits are in the down state coupled with the phonon vacuum. In the strong qubit-photon coupling regime, the photon dynamics can be engineered by tuning the qubit-bath coupling strength $\alpha$ and photon delocalization is achieved by increasing $\alpha$. In the absence of dissipation, photons are localized in the initial resonator. Nevertheless, with moderate qubit-bath coupling, photons are delocalized with quasiequilibration of the photon population in two resonators at long times. In this case, high frequency bath modes are activated by interacting with depolarized qubits. For strong dissipation, photon delocalization is achieved via frequent photon-hopping within two resonators and the qubits are suppressed in their initial down state.
\end{abstract}
\maketitle

\section{Introduction}

In circuit quantum electrodynamics (QED) systems, superconducting qubits are strongly coupled with microwave photons in resonators or transmission lines~\cite{Wallraff2004, Blais2004, Schmidt2010, Hwang2016, Gu2017}. Since conceived in 2004~\cite{Wallraff2004, Blais2004}, circuit QED architectures have been designed and fabricated as research platforms in quantum computation~\cite{Blais2004, Niskanen2007, Helmer2009, Buluta2009, Buluta2011, Houck2012, Georgescu2014, Noh2016} and quantum information~\cite{You2005, Nielsen2010, You2011, Devoret2013}. Due to high flexibility and tunability, circuit QED devices offer the possibility to simulate light-matter interactions in quantum systems with an integrated circuit~\cite{Wallraff2004, Girvin, Houck2012, Schmidt2013, Noh2016}. Experiments focus on engineering the coupling between a single resonator and a qubit for controllable single-resonator systems~\cite{schuster, hofheinz, bishop}. A key challenge is to carry out quantum simulations of strongly correlated photons of coupled-resonator systems by controlling the inter-resonator photon coupling and device-environment interactions~\cite{majer,gerace}. It gives rise to an interesting phenomenon of photon self-trapping due to the competition between the qubit-photon coupling and the inter-resonator photon hopping, which has been realized in experiment using transmon qubits and two coupled transmission-line resonators~\cite{Raftery2014}. Another QED system composed of two coupled nonlinear resonators has been fabricated for quantum amplification~\cite{Eichler2014}. Described as a Bose-Hubbard dimer, this device can also be used for photon generation~\cite{Leonski2004, Miranowicz2006, Liew2010, Bamba2011, Bamba2011_2}.

Recent theoretical studies model the tunnel-coupled resonators each containing a qubit as a Jaynes-Cumming (JC) dimer, which is the smallest possible coupled-resonator system~\cite{Hartmann2006, ciuti, Raftery2014, Schmidt2010, Hwang2016}. The JC Hamiltonian describes a QED system with weak qubit-photon coupling, which omits the counter-rotating-wave (CRW) interactions between the qubit and the photon mode~\cite{Jaynes1963}. Experimental progress has made it possible to achieve ultra-strong coupling~\cite{Wallraff2004, Niemczyk2010, Forn-Diaz2010, Fedorov2010, Yoshihara2017}, where the qubit-photon coupling strength is comparable to the resonator frequency. In this regime, the JC model is invalid and the CRW terms play a crucial role in systems with strongly correlated photons~\cite{Braak2011, LeBoite2016, LeBoite2017, Wangxin2017, Garziano2015, Garziano2016, Kockum2017}. The quantum Rabi model with the CRW interactions considered is expected to provided different physics~\cite{Rabi1936, Rabi1937}. Beyond the JC dimer, Hwang {\it et al.} studied the phase transition of photons in a Rabi dimer~\cite{Hwang2016}. In experimental realizations, fabricated QED systems suffer from ineluctable dissipation stemming from device-environment interactions. The Markovian Lindblad master equation has been applied to describe the photon and qubit dynamics of the JC dimer with the photon decay and the qubit decoherence taken into consideration~\cite{Schmidt2010, Raftery2014}. It is found that dissipation can favor the photon localization in the initial resonator. Although the phase diagram for the photons in a Rabi dimer has been constructed, dissipation induced effects on the dynamics of the photons and the qubits in a Rabi dimer are still not well-understood~\cite{Hwang2016}.

Since the early application of the Markovian Lindblad master equation to describing the dissipative dynamics in the JC dimer~\cite{Schmidt2010}, increasing attention has been attracted to studying influences of dissipation on various QED systems~\cite{Beaudoin2011, Guo2011, Nissen2012, Coto2015, LeBoite2016, Casteels2017, Nazir2016, LeBoite2017}. Most of these studies are conducted by adopting master equations to capture the photon and the qubit dynamics of the QED systems, where environmental effects are considered in a phenomenological manner. However, the interplay between the QED devices and their surroundings is too complex to be modeled by a few dissipative parameters in the methods based on the Markovian Lindblad master equation~\cite{Beaudoin2011, Kaer2013, Nazir2016}. In addition to being affected by bath induced dissipation, the operation of QED devices can benefit from interactions with their surroundings~\cite{Hohenester2009, Calic2011, Roy2011, Carmele2013, Kaer2013, Kaer2013_PRB, Pirkkalainen2013, Muller2015, Nazir2016, Liu2016, Roy-Choudhury2017, Gustin2017, Hornecker2017}. For instance, Hohenester {\it et al.} observed phonon-assisted transition from quantum dot (QD) excitons to photons in nanocavity~\cite{Hohenester2009}. Recently it has been found that exciton-phonon coupling favors single-photon generation in QD-nanocavity systems~\cite{Muller2015}. Therefore, proper treatments of the system-bath interactions are needed.

\begin{figure}
  \centering
  \includegraphics[scale=0.3]{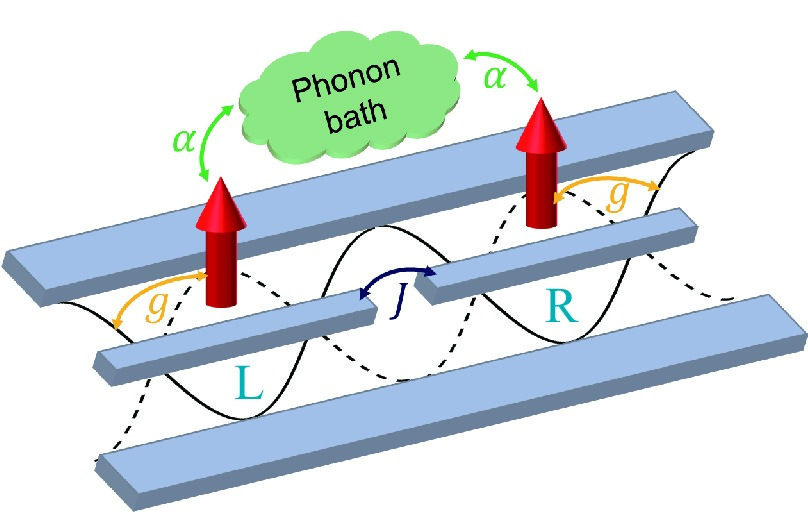}
  \caption{Schematic of the dissipative circuit QED system studied in this work. Photons can hop between two transmission line resonators with a tunneling rate $J$. In each resonator, a qubit interacts with the photon mode with a coupling strength $g$. A phonon bath is used to model the multimode micromechanical resonators and two qubits are coupled to the phonon bath with a strength $\alpha$. }\label{Fig1_sketch}
\end{figure}

Compared with the master equation approach which traces out the bath degrees of freedom (DOFs) in evaluating the reduced density matrices of the target system, the time-dependent variational principle (TDVP) with the Davydov ans\"{a}tze can capture simultaneously the system and the bath dynamics~\cite{Dirac1930,Frenkel1934, Zhao2000_1, Zhao2000_2, Zhao2012, Sun2010, Luo2010}. This approach has been recently extended to include multiple Davydov trial states~\cite{Zhou2015, Zhou2016, Huang2016, Chen2017, Huang2017_2, Wang2017}, producing numerically exact results in a broad parameter regime. Applications have been made to simulate various dynamical processes, such as the dynamics of the Holstein polaron and the spin-boson model~\cite{Sun2010, Luo2010, Zhou2015, Zhou2016, Huang2016, Chen2017, Huang2017_2, Wang2017}, energy transfer in photosynthetic systems~\cite{Ye2012, Huynh2013, Sun2014, Sun2015, Chen2015, Somoza2016, Somoza2017}, singlet fission dynamics~\cite{Fujihashi2017, Huang2017, Fujihashi2017_2}, the Landau-Zener transition~\cite{Huang2018_LZ}, and the qubit and photon dynamics in circuit QED system~\cite{Fujihashi2017_2}.
In addition to the dynamical behavior, the ground state of the Rabi model has been explored with the variational method~\cite{Shore1973_ad2, Chen1989_ad1, Stolze1990_ad3, Hwang2010_ad4}. The trial wave functions were constructed with the displaced oscillator state, the displaced squeezed state, and superpositions of those states. As a superposition of two displaced-squeezed states, a trial state was proposed by Hwang and Choi to capture the squeezing effect of the Rabi model in the ultrastrong coupling regime~\cite{Hwang2010_ad4}.

In this work, combining the TDVP with the multiple Davydov D$_{2}$ ans\"{a}tze, we present a comprehensive study of bath induced effects on the dynamical behavior of a Rabi dimer. Instead of adding dissipation terms into the Lindblad master equation, we model the environmental influences by coupling the qubits to a phonon bath, producing indirect photon-bath interactions. Tuning the qubit-bath coupling strength, modulations on the photon dynamics by the phonon bath can be studied. In addition to the dynamics of the photons and the qubits, the temporal evolution of individual bath modes is depicted explicitly. The reminder of the paper is structured as follows. The Hamiltonian and the methodology applied in this work are described in Section~\ref{sec:Section-II}, including an introduction to the multiple Davydov D$_{2}$ ans\"{a}tze and the TDVP. Observables of interest and parameter configurations are also discussed in this section. Section~\ref{sec:Section-III} documents all the numerical results, covering photon dynamics, qubit polarization and phonon mode populations in various parameter regimes. Concluding remarks are drawn in Section~\ref{sec:Section-IV}.

\section{Model and Methodology \label{sec:Section-II}}

\subsection{Hamiltonian of the hybrid system}

As illustrated in Fig.~\ref{Fig1_sketch}, we consider a dissipative circuit QED device composed of two coupled  transmission line resonators with each interacting with a qubit. The device is modeled as a Rabi dimer and the environmental effects on the device are simulated by coupling the qubits to multimode micromechanical resonators. A phonon bath is adopted to describe the multimode micromechanical resonators. Therefore, the total Hamiltonian for the hybrid system contains three terms
\begin{equation}\label{eq:Htot}
	H=H_{\textrm{RD}}+H_\textrm{B}+H_{\textrm{BQ}}.
\end{equation}

The Rabi dimer can be described by the following Hamiltonian ($\hbar=1$)
\begin{equation}\label{eq:HRD}
	H_{\textrm{RD}}=H_{\textrm{Rabi},\textrm{L}}+H_{\textrm{Rabi},\textrm{R}}-J(a_{\textrm{L}}^{\dagger}a_{\textrm{R}}+a_{\textrm{R}}^{\dagger}a_{\textrm{L}}),
\end{equation}
where $J$ is the photon tunneling amplitude, and
$H_{{\rm Rabi},i}$ ($i=\textrm{L},\textrm{R}$) are the left (L) and right (R) Rabi Hamiltonian, given by~\cite{Rabi1936, Rabi1937, Braak2011, Zhong2017}
\begin{equation}\label{Hrabi}
	H_{\textrm{Rabi},i=\textrm{L}/\textrm{R}} = \frac{\Delta_{i}}{2} \sigma_{z}^{i} + \omega_{i} a_{i}^{\dagger} a_{i} - g_{i} ( a_{i}^{\dagger} + a_{i} ) \sigma_{x}^{i}.
\end{equation}
Here, $\Delta_{i}$ and $\omega_{i}$ are the energy spacing of the qubits and the frequency of the photon mode in the $i$th Rabi system, respectively. $\sigma_{x}^{i}$ and $\sigma_{z}^{i}$ are the usual Pauli matrices, and $a_{i}$ ($a_{i}^{\dagger}$) is the annihilation (creation) operator of the $i$th photon mode. $g_{i}$ characterizes the strength of coupling between the qubits and the photons. In this study, two Rabi sites are assumed to be identical, i.e., $\Delta_{\textrm{L}}=\Delta_{\textrm{R}}=\Delta$, $\omega_{\textrm{L}}=\omega_{\textrm{R}}=\omega_{0}$, and $g_{\textrm{L}}=g_{\textrm{R}}=g$.

The environmental effects on the Rabi dimer are modeled by coupling two qubits to a common phonon bath
\begin{equation}\label{Hb}
	H_\textrm{B}=\sum_{k} \omega_{k} b_{k}^{\dagger} b_{k}
\end{equation}
with an interaction Hamiltonian
\begin{equation}\label{Hbq}
	H_{\textrm{BQ}}=\sum_{k} \phi_{k} (b_{k}^{\dagger}+b_{k})(\sigma_{z}^{\textrm{L}}+\sigma_{z}^{\textrm{R}})
\end{equation}
where $b_{k}$ ($b_{k}^{\dagger}$) is the annihilation (creation) operator of the $k$th bath mode with frequency $\omega_{k}$, and $\phi_{k}$ is the strength of coupling between the $k$th mode and the qubits. The qubit-bath coupling is characterized by
the spectral function,
\begin{equation}
	J(\omega )=\sum_{k} \phi _{k}^{2} \delta( \omega - \omega _{k} )=2 \alpha \omega _{c}^{1-s} \omega ^{s} e^{-\omega/\omega_{c}},
\end{equation}
with $\omega _{c}$ being the cut-off frequency and the dimensionless parameter $\alpha$ quantifying the qubit-bath coupling strength. In our calculations, the photon tunneling $J$ is much smaller than the frequencies of the qubits and the photon modes. As a result of the small $J$, the energy spectrum of the Rabi dimer contains energy levels with small energy gaps. Low frequency bath modes have to be taken into account in our calculations, as these modes may be at resonance with some transitions in the Rabi dimer. It has been demonstrated that the logarithmic discretization procedure is suitable to parameterize the low frequency bath modes with balanced numerical accuracy and efficiency~\cite{WangLu2016}. Therefore, a Sub-Ohmic bath ($s=0.5$) is adopted in this work, and a logarithmic discretization method is used to obtain $\omega_{k}$ and $\phi_{k}$. The cut-off frequency for the bath modes is set to $\omega_{c}=\omega_{0}$, and the maximum frequency used in the discretization is $\omega_{\textrm{max}}=20~\omega_{c}$. To verify our choice of the Sub-Ohmic bath, we have performed test calculations with Ohmic and Super-Ohmic spectral densities. As shown in Fig.~S1 in Supporting Information, it is found that the bath-induced effects on the photon dynamics are independent of the bath types.

\subsection{The multiple Davydov D$_2$ ans\"{a}tze}

The multiple Davydov D$_2$ ans\"{a}tze have been applied to study static and dynamic properties of various systems, producing excellent numerical efficiency and accuracy in a broad parameter regime~\cite{Zhou2015, Zhou2016, Huang2016, Chen2017, Huang2017_2, Wang2017, Fujihashi2017, Huang2017, Fujihashi2017_2, Huang2018_LZ}. In principle, the multiple Davydov D$_2$ ans\"{a}tze can give numerically exact results with a sufficiently high multiplicity. In this study, both the off-diagonal qubit-photon coupling and the diagonal qubit-phonon coupling are included in the system Hamiltonian (\ref{eq:Htot}). Therefore, the multiple Davydov D$_2$ ans\"{a}tze are employed to probe the time evolution of the composite system
\begin{eqnarray}\label{eq:MD2}
	|{\rm D}_{2}^{M}(t)\rangle &=& \sum_{n=1}^{M} \Big[ A_{n} (t)|\uparrow\uparrow\rangle + B_{n} (t) |\uparrow\downarrow\rangle + C_{n} (t) |\downarrow\uparrow\rangle \nonumber\\
	&&~~~~+ D_{n} (t) |\downarrow\downarrow\rangle \Big] \bigotimes |\mu_{n}\rangle_{\textrm{L}}|\nu_{n}\rangle_{\textrm{R}} |\eta_{n}\rangle_{\textrm{B}},
\end{eqnarray}
where $|\uparrow \downarrow \rangle=| \uparrow \rangle_{\textrm{L}} \otimes | \downarrow \rangle_{\textrm{R}}$ with $\uparrow$ $(\downarrow)$ indicating the up (down) state of the qubits. $|\mu_{n}\rangle_{\textrm{L}}$ and $|\nu_{n}\rangle_{\textrm{R}}$ are coherent states of the photon modes
\begin{eqnarray}
	|\mu_{n}\rangle_{\textrm{L}} & = & \exp\left[\mu_{n} (t) a_{\textrm{L}}^{\dagger}-\mu_{n}^{\ast} (t) a_{\textrm{L}}\right]|0\rangle_{\textrm{L}},\\
	|\nu_{n}\rangle_{\textrm{R}} & = & \exp\left[\nu_{n} (t) a_{\textrm{R}}^{\dagger}-\nu_{n}^{\ast} (t) a_{\textrm{R}}\right]|0\rangle_{\textrm{R}},
\end{eqnarray}
where $|0\rangle_{\textrm{L}(\textrm{R})}$ is the vacuum state of the left (right) resonator. $|\eta_{n}\rangle_{\textrm{B}}$ is the coherent state of the phonon bath
\begin{equation}\label{eta}
	|\eta_{n}\rangle_{\textrm{B}} = \exp \left[ \sum_{k}\eta_{nk} (t)  b_{k}^{\dagger}-\eta_{nk}^{\ast} (t) b_{k} \right] |0\rangle_{\textrm{B}}
\end{equation}
with $|0\rangle_{\textrm{B}}$ being the vacuum state of the bath. In Eq.~(\ref{eq:MD2}), $A_{n}(t)$, $B_{n}(t)$, $C_{n}(t)$, $D_{n}(t)$, $\mu_{n}(t)$, $\nu_{n}(t)$, and $\eta_{nk}(t)$ are time-dependent variational parameters to be determined via the TDVP. The physical significance of these variational parameters is clear. For instance, $A_{n}$ is the probability amplitude in the state $|\uparrow\uparrow\rangle|\mu_{n}\rangle_{\textrm{L}}|\nu_{n}\rangle_{\textrm{R}}|\eta_{n}\rangle_{\textrm{B}}$, $\mu_{n}$ ($\nu_{n}$) is the displacement of the left (right) photon mode, and $\eta_{nk}$ is the displacement of the $k$th bath mode.

\subsection{The time-dependent variational principle}

The dynamics of Hamiltonian (\ref{eq:Htot}) is obtained from the Dirac-Frenkel time-dependent variational principle. The equations of motion for all the variational parameters can be derived from
\begin{equation}\label{DiracFrenkel}
 	\frac{d}{dt} \bigg( \frac{\partial L}{ \partial \dot{\alpha}^{*}_{n}} \bigg) - \frac{\partial L}{ \partial \alpha^{*}_{n}} =0.
\end{equation}
Here, $\alpha_{n}$ are the variational parameters, i.e., $A_{n}(t)$, $B_{n}(t)$, $C_{n}(t)$, $D_{n}(t)$, $\mu_{n}(t)$, $\nu_{n}(t)$, and $\eta_{nk}(t)$ in this work. The Lagrangian $L$ is written as
\begin{equation}\label{Lagrangian}
	L = \frac{i}{2} \langle {\rm D}_{2}^{M}(t) | \frac{\overrightarrow{\partial}}{\partial t} - \frac{\overleftarrow{\partial}}{\partial t} | {\rm D}_{2}^{M}(t) \rangle - \langle {\rm D}_{2}^{M}(t) | H | {\rm D}_{2}^{M}(t) \rangle.
\end{equation}

\subsection{Observables}

Combining the TDVP with the multiple Davydov D$_2$ ans\"{a}tze, we are capable of investigating the bath induced dynamics of a Rabi dimer with specific contribution from individual bath modes presented explicitly. In order to study the localization/delocalization of the photons, we calculate the photon numbers in two resonators from expressions
\begin{eqnarray}
	N_{\textrm{L}}(t) & = & \langle{\rm D}_{2}^{M}(t)| a_{\textrm{L}}^{\dagger} a_{\textrm{L}} |{\rm D}_{2}^{M}(t) \rangle \nonumber\\
	& = & \sum_{l,n}^{M} \Big[ A_{l}^{\ast}(t) A_{n}(t)  + B_{l}^{\ast}(t) B_{n}(t)  + C_{l}^{\ast}(t) C_{n}(t) \nonumber\\
	&&~~~~~~~ + D_{l}^{\ast}(t) D_{n}(t) \Big] \mu_{l}^{\ast}(t) \mu_{n}(t)  S_{ln}(t) , \\
	N_{\textrm{R}}(t) & = & \langle{\rm D}_{2}^{M}(t)|a_{\textrm{R}}^{\dagger}a_{\textrm{R}}|{\rm D}_{2}^{M}(t) \rangle \nonumber \\
	& = & \sum_{l,n}^{M} \Big[ A_{l}^{\ast}(t) A_{n}(t)  + B_{l}^{\ast}(t) B_{n}(t)  + C_{l}^{\ast}(t) C_{n}(t) \nonumber\\
	&&~~~~~~~ + D_{l}^{\ast}(t) D_{n}(t) \Big] \nu_{l}^{\ast}(t) \nu_{n}(t)  S_{ln}(t) ,
\end{eqnarray}
where $S_{ln} (t)$ is the Debye-Weller factor
\begin{eqnarray}
	S_{ln} & = &  \exp\left[\mu_{l}^{\ast}(t) \mu_{n}(t)-\frac{1}{2} |\mu_{l}(t)|^2-\frac{1}{2}|\mu_{n}(t)|^2\right] \cdot \nonumber\\
	&& \exp\left[ \nu_{l}^{\ast}(t) \nu_{n}(t) - \frac{1}{2} |\nu_{l}(t)|^2 - \frac{1}{2} |\nu_{n}(t)|^2 \right] \cdot \nonumber\\
	&&\exp \left[ \sum_{k} \Big( \eta_{lk}^{\ast}(t) \eta_{nk}(t) - \frac{1}{2} |\eta_{lk}(t)|^2 - \frac{1}{2} |\eta_{nk}(t)|^2 \Big) \right]. \nonumber\\
\end{eqnarray}
The photon imbalance is then calculated as $Z(t)=N_{\textrm{L}}(t)-N_{\textrm{R}}(t)$ and the normalized photon imbalance is $\tilde{Z}(t)=Z(t)/\big(N_{\textrm{L}}(t)+N_{\textrm{R}}(t)\big)$. These quantities are used to characterize the localization and delocalization of the photons.

In addition to photon dynamics, the time evolution of the qubit states is recorded during the simulations by measuring the qubit polarization via
\begin{eqnarray}
	\langle\sigma_{z}^{\textrm{L}}(t)\rangle & = & \langle{\rm D}_{2}^{M}(t)|\sigma_{z}^{\textrm{L}}|{\rm D}_{2}^{M}(t) \rangle \nonumber \\
	& = & \sum_{l,n}^{M} \Big[ A_{l}^{\ast}(t)A_{n}(t) + B_{l}^{\ast}(t)B_{n}(t) \nonumber\\
	&&- C_{l}^{\ast}(t)C_{n}(t) - D_{l}^{\ast}(t)D_{n}(t) \Big] S_{ln}(t),\\
	\langle\sigma_{z}^{\textrm{R}}(t)\rangle & = & \langle{\rm D}_{2}^{M}(t)|\sigma_{z}^{\textrm{R}}|{\rm D}_{2}^{M}(t) \rangle \nonumber \\
	& = & \sum_{l,n}^{M} \Big[ A_{l}^{\ast}(t)A_{n}(t) - B_{l}^{\ast}(t)B_{n}(t) \nonumber\\
	&&+ C_{l}^{\ast}(t)C_{n}(t) - D_{l}^{\ast}(t)D_{n}(t) \Big] S_{ln}(t).
\end{eqnarray}
As given in Hamiltonian (\ref{eq:Htot}) and illustrated in Fig.~\ref{Fig1_sketch}, the qubits serve as the bridge to connect the photon and the phonon modes, transmitting the bath induced impacts to the photons. Combining influences from the photons and the bath, our calculated qubit dynamics reflects the complex interactions between the photon modes and the phonon bath.

Thanks to the methodology adopted here, the temporal evolution of the phonon bath can also be obtained explicitly. To reveal the participation of individual phonon modes in the Rabi dimer dynamics, we calculate the population on the $k$th mode as follows
\begin{eqnarray}
	N_{k}^{\textrm{B}}(t) &=& \langle{\rm D}_{2}^{M}(t)| b_{k}^{\dagger} b_{k}|{\rm D}_{2}^{M}(t) \rangle \nonumber \\
	&=& \sum_{l,n}^{M} \Big[ A_{l}^{\ast}(t) A_{n}(t) + B_{l}^{\ast}(t) B_{n}(t) + C_{l}^{\ast}(t) C_{n}(t) \nonumber\\
	&&~~~~~~~~+ D_{l}^{\ast}(t) D_{n}(t) \Big] \eta_{lk}^{\ast}(t) \eta_{nk}(t) S_{ln}(t).
\end{eqnarray}
Through interacting with the qubits, the phonon bath gradually gains sufficient energy from the Rabi dimer to affect the dynamics of the photons and the qubits. In return, the influences of the QED system on the bath modes can be investigated by calculating the populations dynamics $N_{k}^{\textrm{B}}(t)$.

\begin{figure}
  \centering
  \includegraphics[scale=0.35]{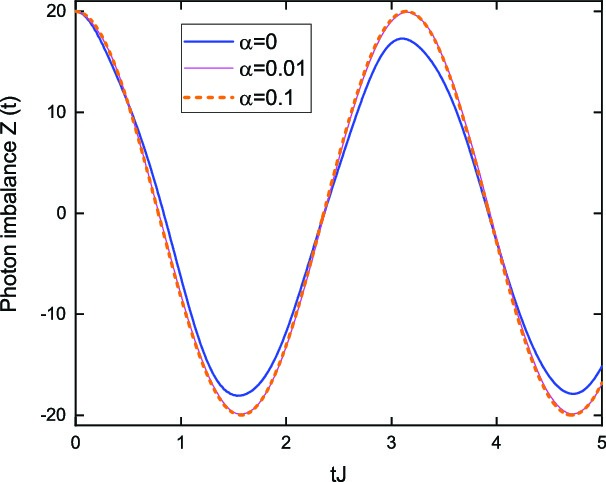}
  \caption{Time evolution of the photon imbalance $Z (t)$ with the qubits coupled to the phonon bath with three coupling strengths $\alpha=0, 0.01 \omega_{0}, 0.1 \omega_{0}$. The qubit-photon coupling strength is $g=0.01~\omega_{0}$ and the inter-resonator photon hopping rate is $J=0.01~\omega_{0}$. Sixty phonon modes are considered in the calculations.}\label{Fig_2_ZandNLR_g001}
\end{figure}

\subsection{Parameter configurations and initial conditions}

In this study, we discuss the case in which the qubits and the photons are at resonance ($\Delta /\omega _{0}=1$). In contrast to the JC dimer, the photon phase diagram in a Rabi dimer is more complicated, and is determined by the competitive effects of $J$ and $g$. For a bare Rabi dimer, a critical photon tunneling rate $J_{c}\approx 0.03~\omega_{0}$ has been proposed~\cite{Hwang2016}, above which the photons are always delocalized regardless of the qubit-photon coupling strength $g$. If $J<J_{c}$, the photon dynamics in a Rabi dimer is found to undergo double phase transitions as the qubit-photon coupling strength $g$ increases~\cite{Hwang2016}. The first transition is from a delocalized phase to a localized phase, then the second one takes place from the localized phase to a new delocalized phase. Photons hop between two resonators in the first delocalized phase, while photons are quasi-equilibrated over two resonators in the second delocalized phase. In order to elucidate the phonon-bath induced effects on the photon dynamics in the Rabi dimer, we choose three combinations of $J$ and $g$, and perform simulations for each parameter configuration with varying qubit-bath coupling strength $\alpha$. In Case \uppercase\expandafter{\romannumeral1}, both $J$ and $g$ are assigned with a value of $0.01~\omega_{0}$, yielding photon delocalization over two resonators in a bare Rabi dimer~\cite{Hwang2016}. In Case \uppercase\expandafter{\romannumeral2}, the photon tunneling amplitude $J$ is set to $0.02~\omega_{0}$ and the qubit-photon coupling strength $g=0.3~\omega_{0}$. The photons in a Rabi dimer are found to be trapped in the initial resonator in this parameter scheme~\cite{Hwang2016}. In Case \uppercase\expandafter{\romannumeral3}, we parameterize the system with $J=0.05~\omega_{0}$ and $g=0.3~\omega_{0}$. As $J > J_{c}$ in this situation, a photon delocalization phase is found for a Rabi dimer~\cite{Hwang2016}. As adopted in common experiments~\cite{Raftery2014}, the photon tunneling rate $J$ used in our work is relatively weak. Therefore, the quadrature-quadrature coupling between the two photon modes~\cite{Rossatto2016, WangYM2016} is neglected in our model, as shown in Eq.~(\ref{eq:HRD}).

Initially, a fully localized photon state is prepared by pumping $N(0)=20$ photons into the left resonator while keeping the right one in a photon vacuum. Experimently it is nontrivial to prepare such an initial photon state. Raftery and coworkers have accomplished such a state through three steps~\cite{Raftery2014}. First, the qubits are detuned by fast flux pulses, turning off the photon-photon interactions. Then an initialization pulse is used to populate the linear resonator modes. Finally, after a variable time delay, the qubits are biased into resonance and the photon-photon interaction is turned on. Along these procedures, one can prepare an initial state with any desired photon imbalance~\cite{Raftery2014}. In contrast to the photons, the qubits in the two resonators start to evolve from their down states and the phonon bath is initially in a vacuum state.

Our calculations are numerically robust regarding the model parameters and it has been verified by preliminary calculations where small deviations were added to the parameters [see Fig.~S2 in Supporting Information]. Although the photon behavior is dependent on the initial photon number, the bath-induced effects on the photon dynamics can be captured by our approach for the case with smaller initial photon number  [see Fig.~S3 in Supporting Information for the calculations with $N(0)=5$]. The numerical convergence also has been carefully checked in preliminary calculations with the sample results shown in Fig.~S3 in Supporting Information.

\section{Results and discussion} \label{sec:Section-III}

With the qubits coupled to a phonon bath, the dynamics of a Rabi dimer will be modulated by the phonon bath in various parameter regimes. Dynamics results calculated by our TDVP approach are presented in this section. In addition to the photon propagation and the qubit polarization, dynamics of individual phonon modes is also explicitly examined.

\subsection{Case \uppercase\expandafter{\romannumeral1}: Weak qubit-photon coupling with weak photon tunneling} \label{SecIIIA} %($J=0.01 ~\omega_{0}, g=0.01~\omega_{0}$)

\begin{figure}%[!htb]
  \centering
  \includegraphics[scale=0.35]{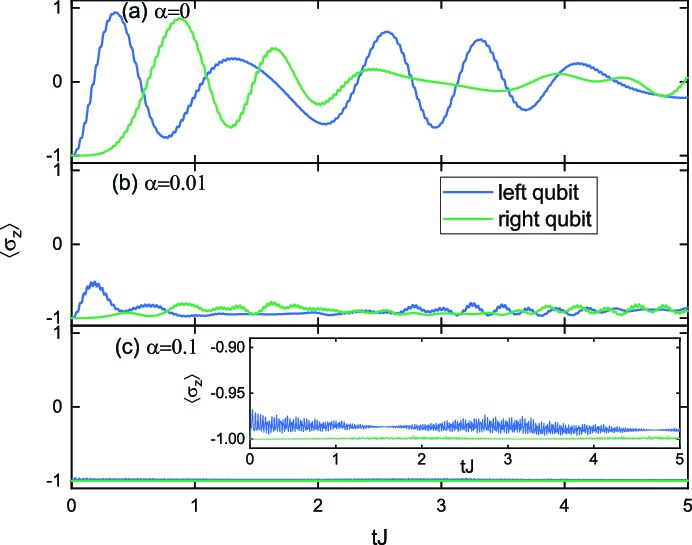}
  \caption{Effects of the qubit-bath coupling on the qubit polarization with $g=0.01~\omega_{0}$ and $J=0.01~\omega_{0}$. Sixty phonon modes are considered in the calculations.} \label{Fig5_sigmaz_g001}
\end{figure}

If the qubit-photon coupling is weak, the photon dynamics is mainly determined by the photon tunneling rate. With a $J$ comparable to $g$, the photons hop between two resonators, yielding photon delocalization~\cite{Hwang2016}. Aiming to study the bath effects on the delocalized photon phase in a Rabi dimer, we first investigate the system dynamics with a weak inter-resonator photon tunneling rate $J=0.01~\omega_{0}$ and a weak qubit-photon coupling strength $g=0.01~\omega_{0}$. The photon imbalance $Z (t)$ is shown in Fig.~\ref{Fig_2_ZandNLR_g001} for three qubit-bath coupling strengths ($\alpha=0, 0.01, 0.1$). The photons of a bare Rabi dimer ($\alpha=0$) are delocalized over two resonators, producing Josephson oscillations of $Z (t)$~\cite{Hwang2016, Guo2011, Larson2011}. If the qubits are weakly coupled to the bath ($\alpha\neq0$), there are no remarkable bath-induced influences on the photon dynamics, as the photons cannot ``feel" the indirect impact from the phonon bath with weak qubit-photon coupling. The photon dynamics is mainly determined by the photon tunneling rate $J$, yielding a Josephson oscillation period of $T_{J}=2\pi/2J$.

\begin{figure}
  \centering
  \includegraphics[scale=0.24]{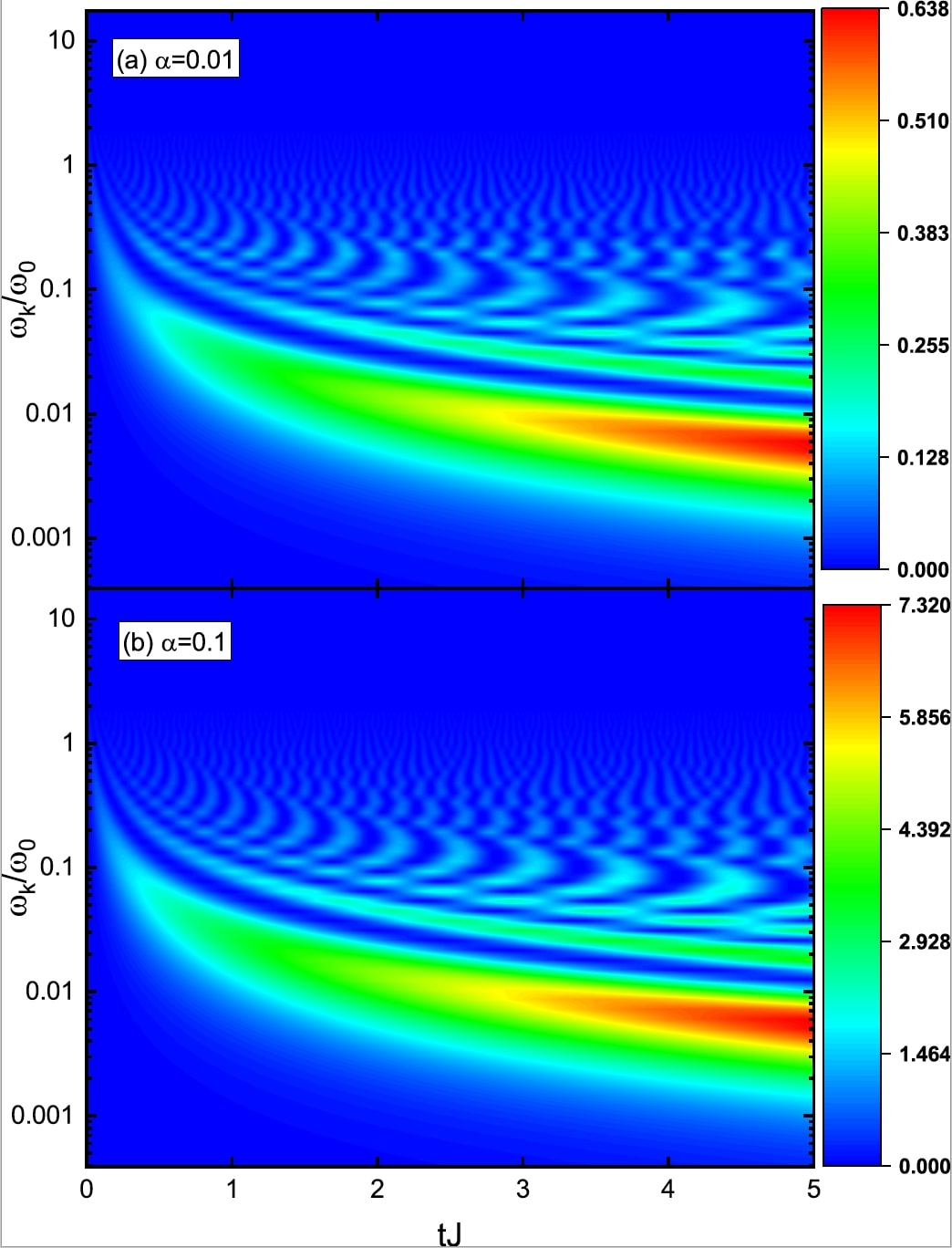}
  \caption{Population dynamics of the bath modes with the qubit-bath coupling $\alpha$ being 0.01 (a), and 0.1 (b). The qubit-photon coupling strength is $g=0.01~\omega_{0}$ and the inter-resonator photon hopping rate is $J=0.01~\omega_{0}$. Sixty phonon modes are considered in the calculations.} \label{Fig8_Popbath_g001}
\end{figure}

In addition to the photon number, the qubit state is another often measured quantity~\cite{Schmidt2010}.
Compared with the photon dynamics, which is almost unaffected by indirect coupling to the phonon bath, the populations of the qubit states are dramatically modulated by the phonon bath, as the qubits are directly coupled to the bath. The time evolution of the qubit polarization is depicted in Fig.~\ref{Fig5_sigmaz_g001} for qubit-bath coupling strengths $\alpha=0$, $0.01$ and $0.1$. In a bare Rabi dimer, two qubits can be excited from the down to the up state via the interaction with the delocalized photons. If all photons are transferred to one resonator, qubit flipping is accelerated in this resonator and is slowed down in the other. Similar phenomenon has been observed in a JC dimer in a semiclassical approximation~\cite{Schmidt2010}. As can be seen from the Rabi Hamiltonian (\ref{Hrabi}), a considerable amount of photons are required to flip the qubits if the qubit-photon coupling is weak. Once the qubit-bath coupling is switched on, oscillations of qubit polarization are greatly suppressed and the two qubits are constrained in the down state, which is similar to the frozen spin in a two-level dissipative system~\cite{Zhang2010}. The stronger the qubit-bath coupling is, the more severe the confinement is. This phenomenon can be explained by the quantum Zeno effect. Conventionally the quantum Zeno effect refers to the suppression of quantum evolution by frequent measurements~\cite{Koshino2005, Harrington2017}. A recent scheme treats the system-environment interactions as ``quasi-measurements"~\cite{Ai2013} which also can lead to quantum Zeno effect~\cite{Harrington2017}. In our model, the qubit-bath coupling works as the quasi-measurements of the state of the qubits, hindering qubit flipping.
The frozen qubits weaken the effective qubit-photon interaction, contributing to decoupling of the photons from the qubits.

\begin{figure}
  \centering
  \includegraphics[scale=0.3]{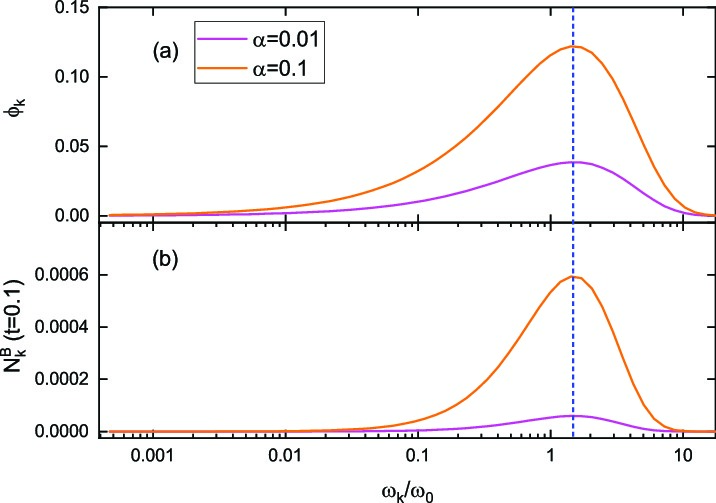}
  \caption{(a) Specific coupling strength $\phi_{k}$ between the $k$th phonon mode and the qubits. (b) Population of individual bath modes at $t=0.1$. As illustrated by the blue vertical dashed line, stronger coupling between the bath mode and the qubits leads to higher population of the corresponding mode in the short time region. The qubit-photon coupling strength is $g=0.01~\omega_{0}$ and the inter-resonator photon hopping rate is $J=0.01~\omega_{0}$. Sixty phonon modes are considered in the calculations.}\label{Fig_5_PhikPopBath_g001}
\end{figure}

In contrast to the density matrix methods that trace out the DOFs of the bath, the TDVP with the Davydov ans\"{a}tze can treat the DOFs of the qubits, the photons and the bath modes explicitly, making it possible to study the time evolution of each bath mode. In order to investigate the participation of individual bath modes in the Rabi dimer dynamics, we calculate the population of the bath modes, and the results with $g=0.01~\omega_{0}$ and $J=0.01~\omega_{0}$ are shown in Fig.~\ref{Fig8_Popbath_g001}. The shape of the population distribution over all modes is independent of the qubit-bath coupling, while the population magnitude increases with $\alpha$. For short times, the bath mode population is associated with the strength of coupling between specific bath mode and the qubits. As shown in Fig.~\ref{Fig_5_PhikPopBath_g001}, the phonon modes with frequencies near $\omega_{k} = 1.5~\omega_{0}$ have relatively strong coupling to the qubits, giving rise to high populations of these modes at short times. Within this short time interval, there is energy flowing into the phonon bath. As the time evolves, it is difficult to transfer more energy to the bath as the qubits cannot flip. It follows that the energy within the bath redistributes by gradually populating the low-frequency modes.

\begin{figure}
  \centering
  \includegraphics[scale=0.32]{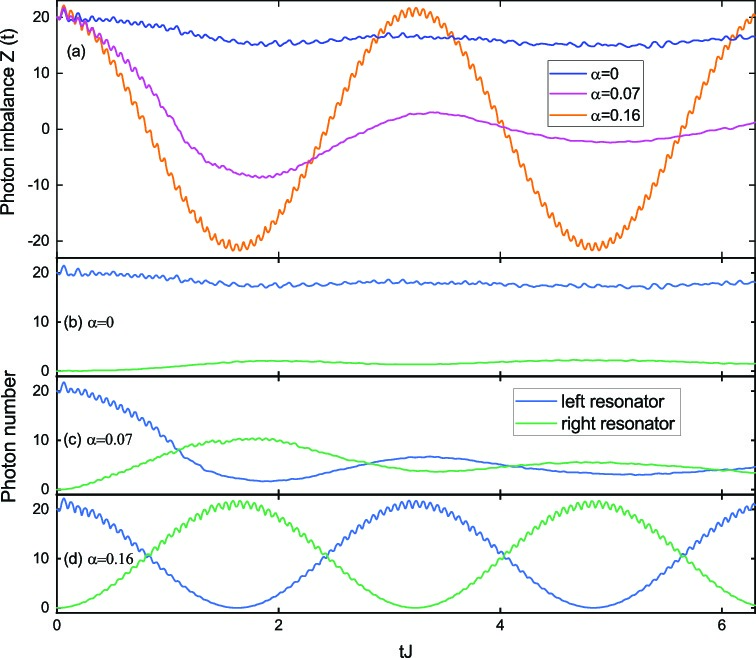}
  \caption{ (a) Effects of the qubit-bath coupling $\alpha$ on the time evolution of the photon imbalance $Z$. (b)-(d) Time evolution of the photon numbers in the left ($N_{\textrm{L}}$) and right ($N_{\textrm{R}}$) resonators with different qubit-bath coupling $\alpha$. The qubit-photon coupling strength is $g=0.3~\omega_{0}$ and the inter-resonator photon hopping rate is $J=0.02~\omega_{0}$. Sixty phonon modes are considered in the calculations.}\label{Fig_3_ZandNLR_g02}
\end{figure}

\subsection{Case \uppercase\expandafter{\romannumeral2}: Strong qubit-photon coupling with weak photon tunneling} %($J=0.02 ~\omega_{0}, g=0.3~\omega_{0}$)

By increasing the qubit-photon coupling, Hwang and coworkers have shown that the photons undergo a transition from a delocalized phase to a localized one if the photon tunneling amplitude is weak~\cite{Hwang2016}. Configuring the Rabi dimer with $J=0.02 ~\omega_{0}$ and $g=0.3~\omega_{0}$, we study the bath induced effects on the photon dynamics and the results are presented in Fig.~\ref{Fig_3_ZandNLR_g02}. In this parameter regime, the photons are localized in the left resonator of a bare Rabi dimer ($\alpha=0$), a result of the qubit-photon interactions. With non-negligible $g$, the photon dynamics is dominated by the nonlinearity of the Rabi dimer spectrum, which prevents the photons from hopping to the right resonator. This phenomenon is known as the photon-blockade effect \cite{Imamoglu1997, Birnbaum2006, Lang2011, Ridolfo2012, Hwang2016, LeBoite2016}. Once the dissipation is taken into account by switching on the qubit-bath coupling $\alpha$, the photons are no longer localized in the initial resonator.

For a qubit-bath coupling strength of $\alpha=0.07$, the photons are found to escape from the left resonator through two channels. On one hand, the photons tunnel to the right resonator at short times. On the other hand, some photons are dissipated due to the coupling to the phonon bath via the qubits, producing a gradually decreasing total photon number. At long times, only a small portion of photons are conserved and delocalized in two resonators. The photon delocalization is characterized by quasiequilibration of the photon population with two resonators having almost the same number of photons. Therefore, the photon imbalance oscillates around zero with decreasing amplitudes as time evolves. With stronger coupling, e.g., $\alpha=0.16$, there is almost no loss of the total photon number, since the indirect photon-bath coupling is weakened by the qubits frozen in the down state. Surprisingly, the photons are delocalized via frequent hopping within two resonators, which is similar to that for lower values of $J$ and $g$, e.g., $J=0.01~\omega_{0}$ and $g=0.01~\omega_{0}$ [see Fig.~\ref{Fig_2_ZandNLR_g001}]. Comparing the photon localization for $\alpha=0$ and the photon delocalization for $\alpha=0.16$, we find that the photon confinement due to the qubit-photon coupling can be eliminated by strong qubit-bath coupling. From this set of calculations, we can clearly see that the dissipative bath induces two different forms of photon delocalization in the strong coupling regime, which is controlled by tuning the qubit-bath coupling strength $\alpha$. For instance, a moderate $\alpha$ can be applied to achieve the distinguished photon delocalization with quasiequilibration of the photons, which is usually observed in the deep-strong coupling regime without dissipation~\cite{Hwang2016}. If the qubit-bath coupling is strong, the total photon number is conserved and the photons are delocalized over two resonators via hopping.

\begin{figure}%[!htb]
  \centering
  \includegraphics[scale=0.32]{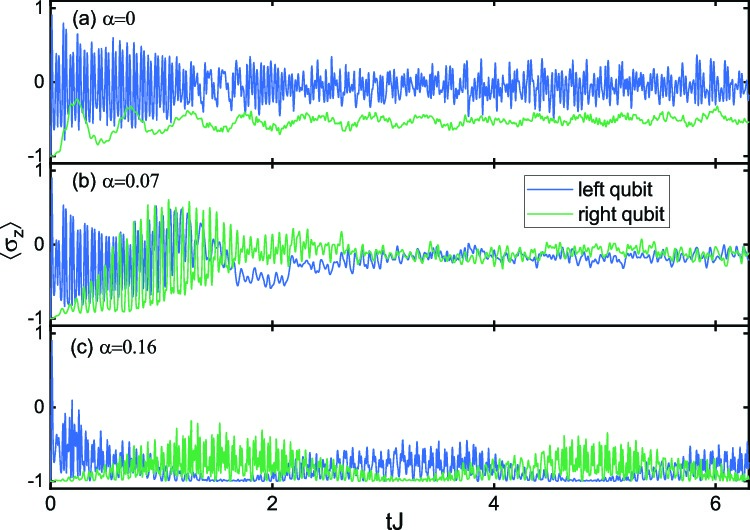}
  \caption{Effects of the qubit-bath coupling on the qubit polarizations with $g=0.3~\omega_{0}$ and $J=0.02~\omega_{0}$. Sixty phonon modes are considered in the calculations.} \label{Fig6_sigmaz_g02}
\end{figure}

In order to understand how the phonon bath impacts the photon dynamics, we study the qubit dynamics by calculating the qubit polarization. Results are collected in Fig.~\ref{Fig6_sigmaz_g02}. For a Rabi dimer without the phonon bath, the photons are trapped in the left resonator and help flip the left qubit frequently, as shown in Fig.~\ref{Fig6_sigmaz_g02} (a). In contrast to the left qubit, the polarization of the right qubit oscillates around the value of $-\frac{1}{2}$, indicating that the right qubit tends to reside in its down state throughout.

\begin{figure}
  \centering
  \includegraphics[scale=0.24]{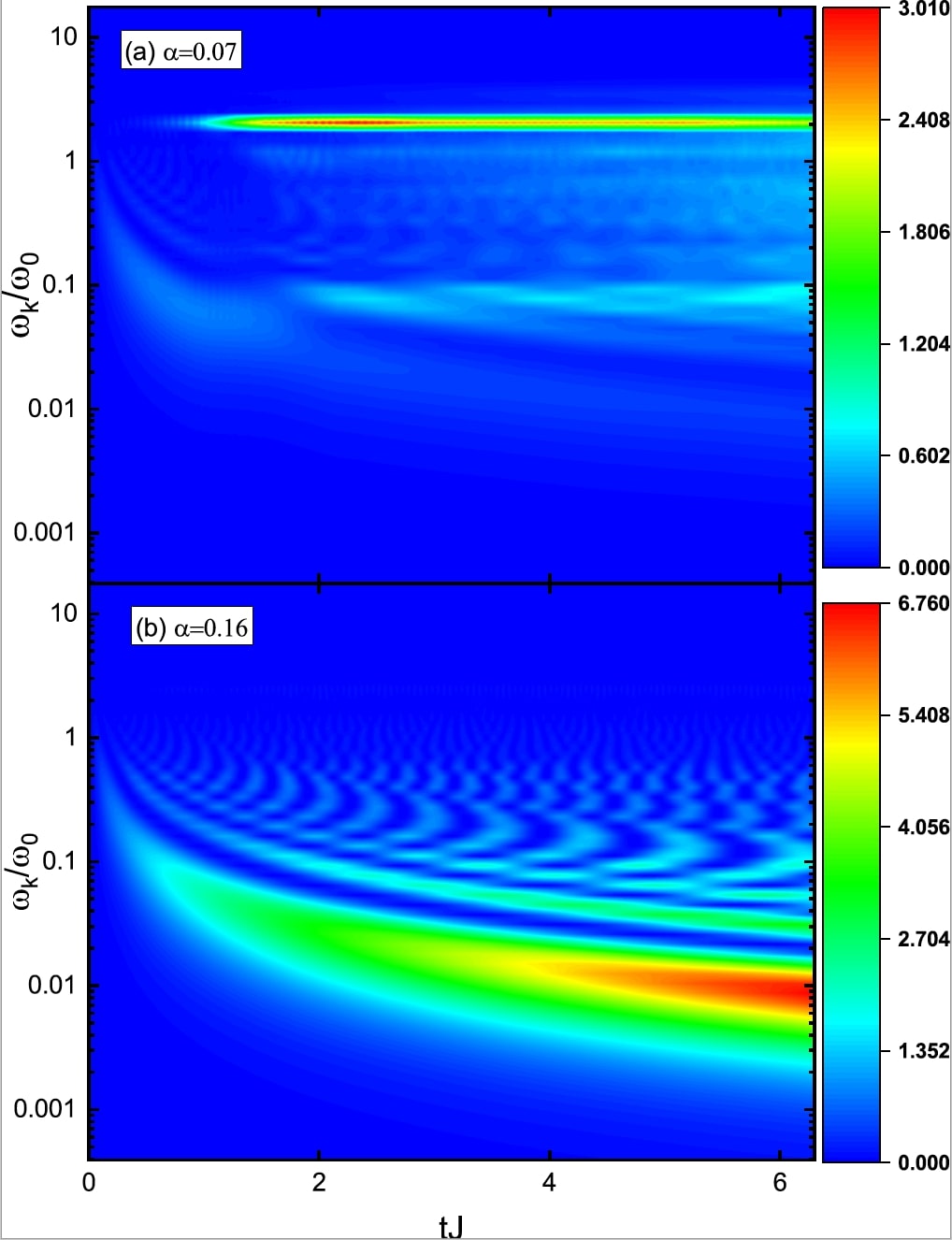}
  \caption{Population dynamics of the bath modes with the qubit-bath coupling $\alpha$ being 0.07 (a), and 0.16 (b). The qubit-photon coupling strength is $g=0.3~\omega_{0}$ and the inter-resonator photon hopping rate is $J=0.02~\omega_{0}$. Sixty phonon modes are considered in the calculations.} \label{Fig9_Popbath_g02}
\end{figure}

If the qubit-bath coupling is moderate ($\alpha=0.07$), the right qubit stays in the down state at short times ($tJ<0.6$). In this regime, the flipping of the left qubit is quenched, promoting the photon delocalization [see Fig.~\ref{Fig_3_ZandNLR_g02} (c)]. Stimulated by the photons flowing to the right resonator, the right qubit gradually acquires sufficient energy to flip between its down and up states. In the time interval of $0.6< tJ < 2.0$, two qubits frequently flip, and a bridge between the photons and the phonon bath is established, producing a rapid decay of the total photon number and a sharp increase of the bath mode population [see Fig.~\ref{Fig9_Popbath_g02} (a)]. At long times ($tJ>2.0$), two qubits are depolarized with $\sigma_z^{\textrm{L}(\textrm{R})}\sim0$, which is useful to characterize the photon delocalization with quasiequilibration of photons in two resonators~\cite{Hwang2016}.

If the qubit-bath coupling is stronger, e.g., $\alpha=0.16$, two qubits stay in the down states most of the time. This phenomenon originates from the fact that the diagonal qubit-bath coupling considerably increases the effective bias of the qubits and removes the qubit-photon resonance, preventing qubit flipping. Through the off-diagonal qubit-photon coupling [see Eq.~(\ref{Hrabi})], localized photons stimulate qubit flipping, which can help trap the photons in return. As the two qubits are confined in the down state, the effects of the qubit-photon interactions on the photon dynamics can be eliminated. It seems that the photon modes are decoupled from the subsystem composed of the qubits and the phonon bath. Therefore, the photon hopping between the two resonators is mainly due to the photon tunneling. Bridging the photons and the bath modes, the qubit can tune the indirect coupling between the photons and the bath via qubit flipping. If the qubit is localized in one state, bath induced effects cannot be experienced by the photons. Hence, there is no detectable decay of the total photon number in this scenario, as is shown in Fig.~\ref{Fig_3_ZandNLR_g02} (d). At short times, the left qubit flips, while the right qubit stays in the down state [see Fig.~\ref{Fig6_sigmaz_g02} (c)]. In this time interval, the photons are completely localized in the left resonator while helping flip the left qubit. In contrast, the phonon bath is initiated from a vacuum state and cannot confine the left qubit in its down state within a short time.

In order to elucidate contributions of individual bath modes to the manipulation of the photon and qubit dynamics in the Rabi dimer, we calculate the bath mode populations, and the results are illustrated in Fig.~\ref{Fig9_Popbath_g02}. In the moderate qubit-bath coupling case ($\alpha=0.07$), several low frequency bath modes get excited with small populations in short times ($tJ<0.6$). In the interval of $0.6<tJ<2.0$, the modes with frequency around $\omega_{k}=2.0~\omega_{0}$ are promptly populated. Within this period, two qubits can flip, effectively turning on the indirect photon-bath coupling and leading to a rapid reduction of the total photon number.
Increasing population in the phonon modes with $\omega_{k}\sim2.0~\omega_{0}$ is contributed from the energy transferred by the two qubits from the photons to the phonon modes. Similar phenomenon has been reported recently by Garziano and coworkers in a model where a photon mode is coupled to two separate atoms~\cite{Garziano2015, Garziano2016}. When the photon frequency is twice the atomic transition frequency, they found that the two atoms can be simultaneously excited by the photon, and this process is reversible~\cite{Garziano2015, Garziano2016}. In our study, the phonon modes with $\omega_{k}\sim2.0~\omega_{0}$ are populated through coupling to two qubits, analogous to the reversible process demonstrated in Ref.~\cite{Garziano2016}, i.e., two atoms can jointly emit a single photon during downward transitions from their excited states.
With the passing of time, more bath modes are excited with frequencies ranging from $0.05~\omega_{0}$ to $2.0~\omega_{0}$. If the qubit-bath coupling is strong, i.e., $\alpha=0.16$, the bath-induced effects are too weak to impact the photons as the qubits are confined in the down state. Therefore, the phonon dynamics is similar to the weak qubit-photon coupling case (see Fig.~\ref{Fig8_Popbath_g001}), where the population gradually flows to the low frequency modes.

From aforementioned results and discussions, it is found that the photon dynamics in a Rabi dimer can be controlled by manipulating the qubit state. Schmidt and coworkers demonstrated that the photon phase transition can be detected by measuring the qubit state~\cite{Schmidt2010}. Recently, Baust {\it et al.} have achieved tunable coupling between two resonators by controlling the state of the qubit that connects the resonators~\cite{Baust2015}. In their experiment, the qubit population is controlled by a microwave drive~\cite{Baust2015}.
Nori and coworkers have proposed a multioutput single-photon device by coupling two resonators to a qubit~\cite{Wang2016}. They have also studied the phonon blockade in nanomechanical resonators coupled to a qubit~\cite{Liu2010, Wang2016_2}. Some optomechanical systems have been conceived with the photon mode, the mechanical resonator, and the two-level system coupled together~\cite{Restrepo2017, Restrepo2014, Holz2015, Akram2015, ZhouBY2016}.
Pirkkalainen and coworkers have fabricated a QED device coupled with phonons, in which a superconducting transmon qubit is coupled to a microwave cavity and a micromechanical resonator~\cite{Pirkkalainen2013, Pirkkalainen2015}. The qubit-phonon coupling strength also can be measured and regulated~\cite{Pirkkalainen2013, Xiang2013, Xiong2015, Pirkkalainen2015, Hartke2017, Rouxinol2016}. In our study, the qubits in a Rabi dimer are coupled to a multimode phonon bath. By tuning the qubit-bath coupling, the qubit state can impact the photon dynamics via the qubit-photon interaction. Thanks to the advances in nanofabrication, multimode micromechanical resonators can be manufactured to serve as the phonon bath~\cite{Massel2012, Mian2012}. The hybrid QED system proposed in this work is experimentally feasible. Such devices can be fabricated not only to control the photon dynamics in the QED systems as demonstrated here, but also to serve as a platform for fundamental studies of quantum physics, such as the photon-qubit-phonon interactions in QED systems.

\begin{figure}
  \centering
  \includegraphics[scale=0.32]{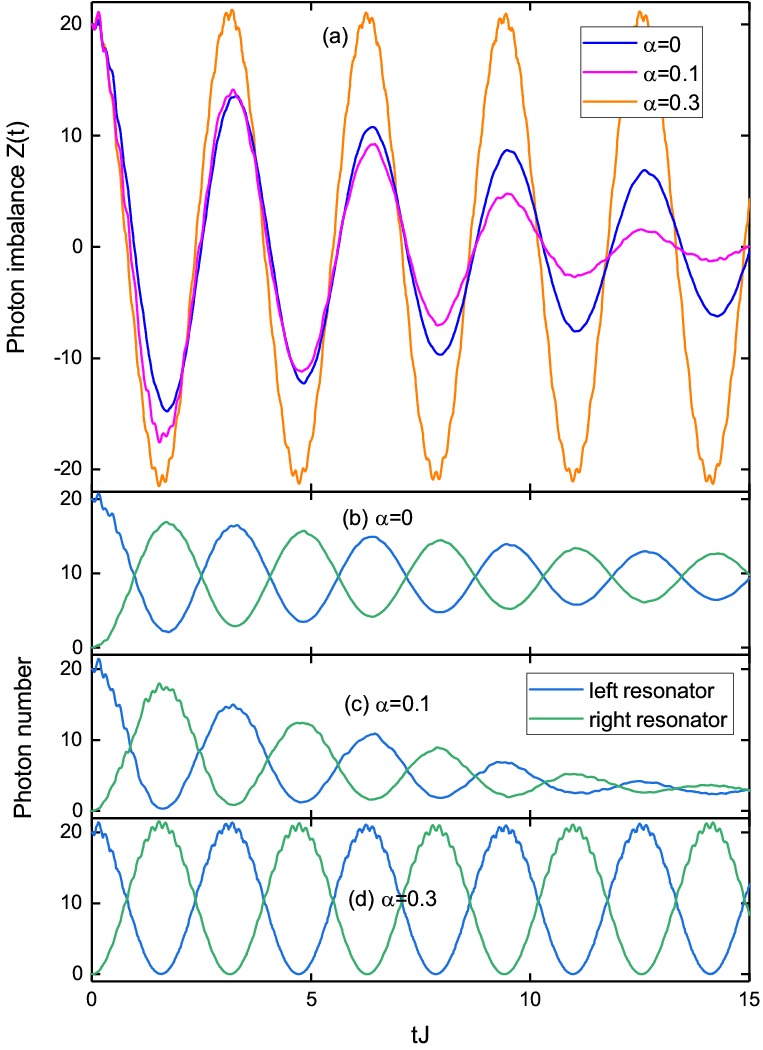}
  \caption{ (a) Effects of the qubit-bath coupling $\alpha$ on the time evolution of the photon imbalance $Z$. (b)-(d) Time evolution of the photon numbers in the left ($N_{\textrm{L}}$) and right ($N_{\textrm{R}}$) resonators with different qubit-bath coupling $\alpha$. The qubit-photon coupling strength is $g=0.3~\omega_{0}$ and the inter-resonator photon hopping rate is $J=0.05~\omega_{0}$. Sixty phonon modes are considered in the calculations.}\label{Fig_9_ZandNLR_J005}
\end{figure}

\begin{figure}%[!htb]
  \centering
  \includegraphics[scale=0.32]{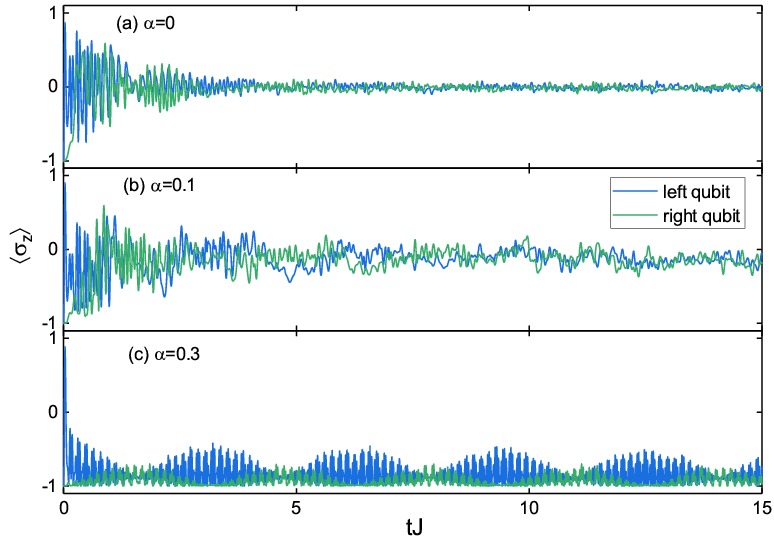}
  \caption{Effects of the qubit-bath coupling on the qubit polarizations with $g=0.3~\omega_{0}$ and $J=0.05~\omega_{0}$. Sixty phonon modes are considered in the calculations.} \label{Fig10_sigmaz_J005}
\end{figure}

\subsection{Case \uppercase\expandafter{\romannumeral3}: Strong qubit-photon coupling with moderate photon tunneling} % ($J=0.05 ~\omega_{0}, g=0.3~\omega_{0}$)

\begin{figure}
  \centering
  \includegraphics[scale=0.24]{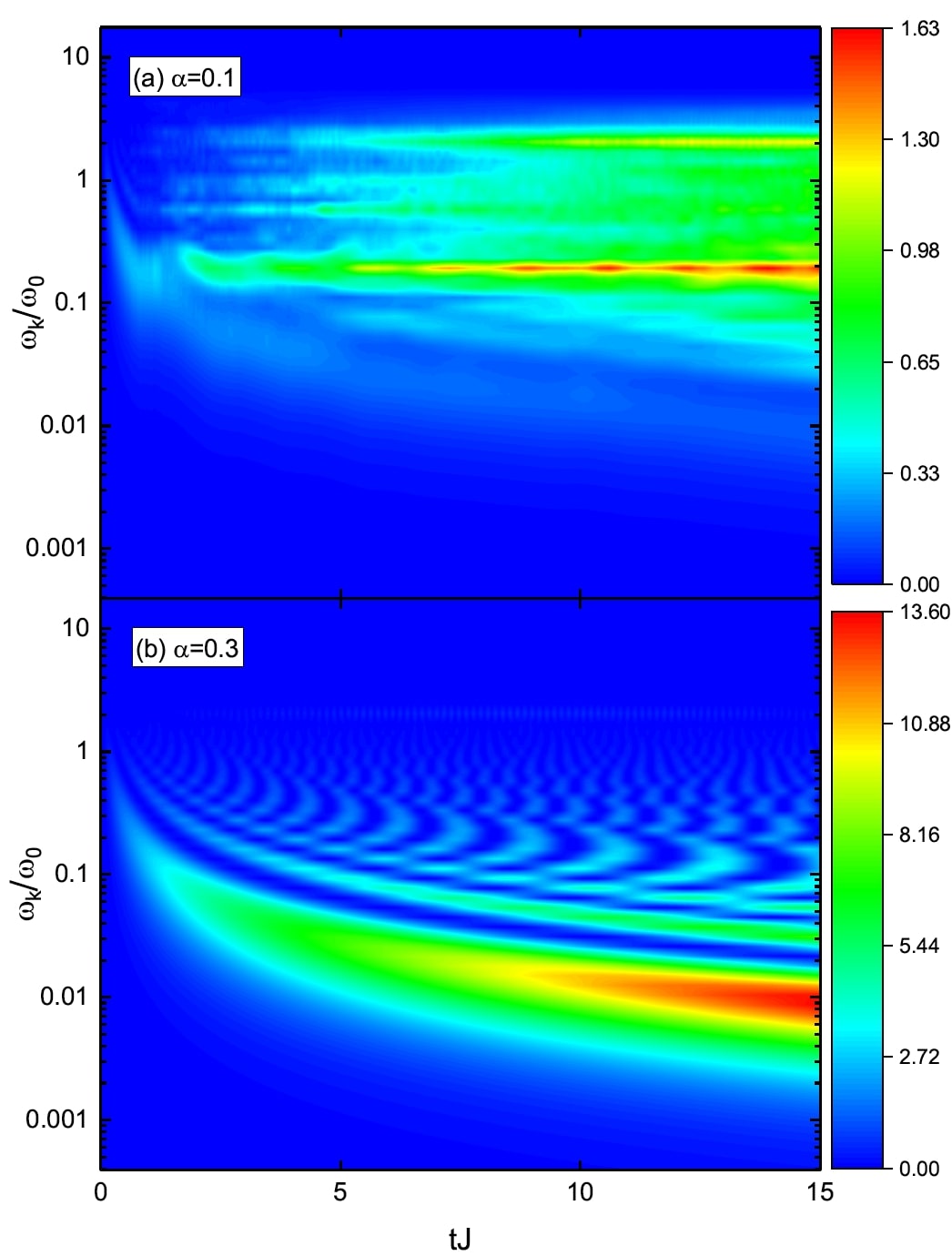}
  \caption{Population dynamics of the bath modes with the qubit-bath coupling $\alpha$ being 0.1 (a), and 0.3 (b). The qubit-photon coupling strength is $g=0.3~\omega_{0}$ and the inter-resonator photon hopping rate is $J=0.05~\omega_{0}$. Sixty phonon modes are considered in the calculations.} \label{Fig11_Popbath_J005}
\end{figure}

It has been reported that the photons are delocalized in the parameter configuration of $J=0.05 ~\omega_{0}$ and $g=0.3~\omega_{0}$ \cite{Hwang2016}. The bath effects on the photon and qubit dynamics in this configuration are presented here. As shown in Fig.~\ref{Fig_9_ZandNLR_J005} (a), the photon imbalance for a bare Rabi dimer oscillates around zero with decreasing amplitudes, indicating photon delocalization with quasiequilibration of photons in two resonators at long times. Coupling the qubits to the phonon bath with a strength of $\alpha=0.1$ accelerates the decay of the total photon number, although the photon imbalance is similar to that of the bare Rabi dimer. In this parameter configuration, the two qubits can freely flip [see Fig.~\ref{Fig10_sigmaz_J005} (b)] and the bath induced dissipation is effective on the photons, leading to continuous decrease of the total photon number. Many high frequency modes are activated by interacting with the Rabi dimer [see Fig.~\ref{Fig11_Popbath_J005} (a)].

Increasing the qubit-bath coupling strength $\alpha$ to 0.3, the photons are still delocalized over the two resonators and the decay of the total photon number is greatly decelerated, as can be seen in Fig.~\ref{Fig_9_ZandNLR_J005} (d). In this case, two qubits remain in the down states [see Fig.~\ref{Fig10_sigmaz_J005} (c)] and it is difficult to build indirect photon-phonon coupling. In contrast to the case of $\alpha=0.1$ where many high-frequency phonon modes are excited, the phonon population gradually flows to the low frequency modes with $\alpha=0.3$, as shown in Fig.~\ref{Fig11_Popbath_J005} (b).

\section{Concluding remarks}\label{sec:Section-IV}{}

Equipped with the multiple Davydov D$_2$ ans\"{a}tze, the Dirac-Frenkel time-dependent variational principle is applied to investigate the bath induced effects on the dynamics of a Rabi dimer by coupling the qubits to a common phonon bath. The dynamics of the photons, the qubits and the bath modes are studied in various parameter regimes. Through extensive calculations, it is found that the photon dynamics in a Rabi dimer can be controlled by manipulating the qubit states via tuning the qubit-bath coupling. For weak qubit-photon coupling, the photon dynamics is almost unaffected by the phonon bath, and the photons are delocalized by hopping between two resonators. However, the qubits are frozen in the initial down state due to their strong coupling to the bath. In the strong coupling regime with weak photon tunneling, the photons are localized in the initial resonator in the absence of the dissipation. With inclusion of the environmental influences, the total photon number is reduced, and photons are delocalized in two resonators. At long times, the two resonators have almost the same number of photons, accompanied by the depolarization of the qubits, $\sigma_z^{\textrm{L}(\textrm{R})}\sim0$. For a large dissipation strength $\alpha$, the indirect photon-bath coupling is suppressed by the qubits frozen in their down state. Consequently, the photons can freely hop between the two resonators via the inter-resonator tunneling rate $J$, producing a photon delocalization with oscillating photon imbalance. The two kinds of photon delocalization can be achieved by tuning the qubit-bath interaction amplitudes. These intriguing features are attributed to the environmental effects in the strong qubit-photon coupling regime. It is expected that the hybrid QED device proposed in here can be fabricated in future experiments, which will provide a platform for the studies of quantum physics.

\section*{Supporting Information}
Supporting Information is available from the Wiley Online Library or from the author.

\section*{Acknowledgments}
This work is supported by Singapore National Research Foundation through the Competitive Research Programme (CRP) under Project No. NRF-CRP5-2009-04 and the Singapore Ministry of Education Academic Research Fund Tier 1 (Grant No. RG106/15).

%\appendix
%\section{AAA}

%\bibliographystyle{andp2012}%.bst
%\bibliography{RD_bib}

\providecommand{\WileyBibTextsc}{}
\let\textsc\WileyBibTextsc
\providecommand{\othercit}{}
\providecommand{\jr}[1]{#1}
\providecommand{\etal}{~et~al.}

\end{document}